\pdfoutput=1
\documentclass[aps,pre, twocolumn, groupedaddress]{revtex4-1}
\usepackage{lipsum}
\usepackage{mathtools}
\usepackage{graphicx}
\usepackage{dcolumn}
\usepackage{amssymb}
\usepackage{amsmath}    
\usepackage{bm}
\usepackage{hyperref}
\usepackage{latexsym}
\usepackage{verbatim}
\usepackage{color}
\usepackage{float}
\usepackage[caption=false]{subfig}
\usepackage{epstopdf}

\def\beq{\begin{equation}}
\def\eeq{\end{equation}}

\def\beq{\begin{equation}}                          
\def\eeq{\end{equation}}                          
\def\bea{\begin{eqnarray}}                          
\def\eea{\end{eqnarray}}

\DeclareRobustCommand{\uvec}[1]{{%
  \ifcsname uvec#1\endcsname
     \csname uvec#1\endcsname
   \else
    \bm{\hat{\mathbf{#1}}}%
   \fi
}}

\preprint{}

\begin{document}

\title{Active polar flock with birth and death}

\author{Pawan Kumar Mishra}
\email[]{pawankumarmishra.rs.phy19@itbhu.ac.in}
\affiliation{Department of Physics, Indian Institute of Technology (BHU), Varanasi, U.P. India - 221005}

\author{Shradha Mishra}
\email[]{smishra.phy@itbhu.ac.in}
\affiliation{Department of Physics, Indian Institute of Technology (BHU), Varanasi, U.P. India - 221005}
\date{\today}

\begin{abstract}
	We study a collection of  self-propelled polar particles on a two-dimensional substrate
	with birth and death. 
	We introduce a minimal 
	lattice model for the system using active Ising spins, 
	where each particle  can have two possible 
	orientations. The activity is modeled as a biased movement of the 
	particle along its direction of orientation.
	The particles also  
	align with their nearest neighbors 
	 using Metropolis Monte-Carlo algorithm. 
	System shows a disorder-to-order transition 
	by tuning the temperature of the system. 
	Additionally, the birth and death of the particles is introduced 
	through a birth and death rate
	$\gamma$.  The system is studied near the disorder-to-order 
	transition. The nature of disorder-to-order transition shows a
	crossover from first order, discontinuous  to continuous type 
	as we tune $\gamma$ from zero to finite values. We also write the effective 
	free energy of the local order parameter using renormalised mean field theory
	and it confirms the dependence of the nature of phase transition on the birth and 
	death rate parameter.

\end{abstract}

\maketitle
\section{Introduction}
The active matter systems can be recognised as a collection of particles in which 
the individual components possess non-zero 
motility  by converting the energy from its surroundings
and also from the medium \cite{Toner2005, Toner1995, tonertupre1998, rmp2013Marchetti, rmp2016Bechinger, pr2012Vicsek}. 
The  active particles  spontaneously self organize when present in large numbers, 
and results in coordinated and collective behavior (CB) 
on various length scales  \cite{TopicalReview2020, rmp2013Marchetti, rmp2016Bechinger, Saintillan2010, Shen2004, Dombrowski2004, Kemkemer2000, Surrey2001, Bendix2008}. 
The phenomenon of collective behavior is 
being studied with great interest in systems exhibiting nonequilibrium phase transition under driven noise and particle density \cite{Toner1995, chatepre2008, speckexpprl2013, shradhamanna2016, shradhajocp2018, shradhajstat2021}.
In different studies of active matter systems it has been shown that the system changes 
	its properties such as pattern of the structure, nature of the phase transition by 
	tuning the 
	interaction  among the particles \cite{rmp2013Marchetti, shradhamanna2016, shradhajocp2018, shradhajstat2021, viewanglepre2016, Giomi2010, Ramaswamy2003}. Among them the nature of phase  transition is one 
	of the 
	most studied phenomenon in this field of research \cite{chatepre2008, shradhamanna2016, shradhajocp2018, shradhajstat2021, viewanglepre2016, SolonPRL2013, SolonPRE2015}. 
	The phase transition in the collection of  self-propelled particles also called as ``flocking transition'' is important, 
	because it can be described as 
	 a nonequilibrium analog of disorder-to-order phase transition in equilibrium  systems \cite{Vicsek1995, SolonPRE2015, SolonPRL2013}. \\
In recent study of Solon {\em et. al.} \cite{SolonPRL2013}, it has been found that the nature of phase transition in self-propelling agents 
is  analogous to the liquid-gas transition  on the variation of  temperature and density in the system \cite{SolonPRL2013, SolonPRE2015}. 
	To understand the phase transition  Solon {\em et. al.} introduced a microscopic lattice model with discrete symmetry, which is known as active Ising model $(AIM)$ \cite{SolonPRL2013}. 
	The $AIM$ is much simplified model for the collective motion  and gives the basic features of the flocking models \cite{SolonPRE2015}: viz, band formation, large density fluctuations,
	discontinuous disorder-to-order phase transition etc.
	The study of $AIM$ by \cite{SolonPRL2013} is for the system where total number of agents is fixed. The effect of birth and death of agents on the system is not yet explored. \\
	In this current study we ask the question: whether  the introduction of 
	 birth and death of the agents can affect the nature 
	 of phase transition? To serve this purpose, we introduced a minimal  
	 lattice based-model of active Ising spins $(AIM)$ with an additional birth 
	 and death rate $\gamma$. \\
	The system is studied for 
	various $\gamma$ and it is found that for the $\gamma=0$, the system
	shows a first order, disorder-to-order phase transition with the appearance of bands 
	in the local density and 
	magnetisation. On introducing the $\gamma$, the bands start to dilute and finally disappear for large 
	$\gamma$ and transition becomes continuous in nature. 
	We also studied the system using coarse-grained hydrodynamic 
	equations of motion.  Using renormalised mean field theory we write an 
	effective free energy for local order parameter and find 
	an additional cubic order nonlinearity present for zero birth and death and the nonlinearity weakens on 
	increasing birth and death rate.\\

	The rest of the paper is organized as follows. In Sec. \ref{model}, we discuss the model and simulation details. In Sec. \ref{results}, the results from the numerical simulations are discussed where we mainly conclude how the nature of phase transition is changing by tuning the parameter  $\gamma$. In addition to numerical approach, we also study the system analytically in Sec. \ref{analytics} with the help of coarse-grained hydrodynamic equations for density and polarisation using renormalised mean field theory. Finally in  Sec. \ref{discussion}, we conclude the paper with a summary and discussion of the results.
 
\begin{figure}      

  \includegraphics[width=8.4cm,height=8cm]{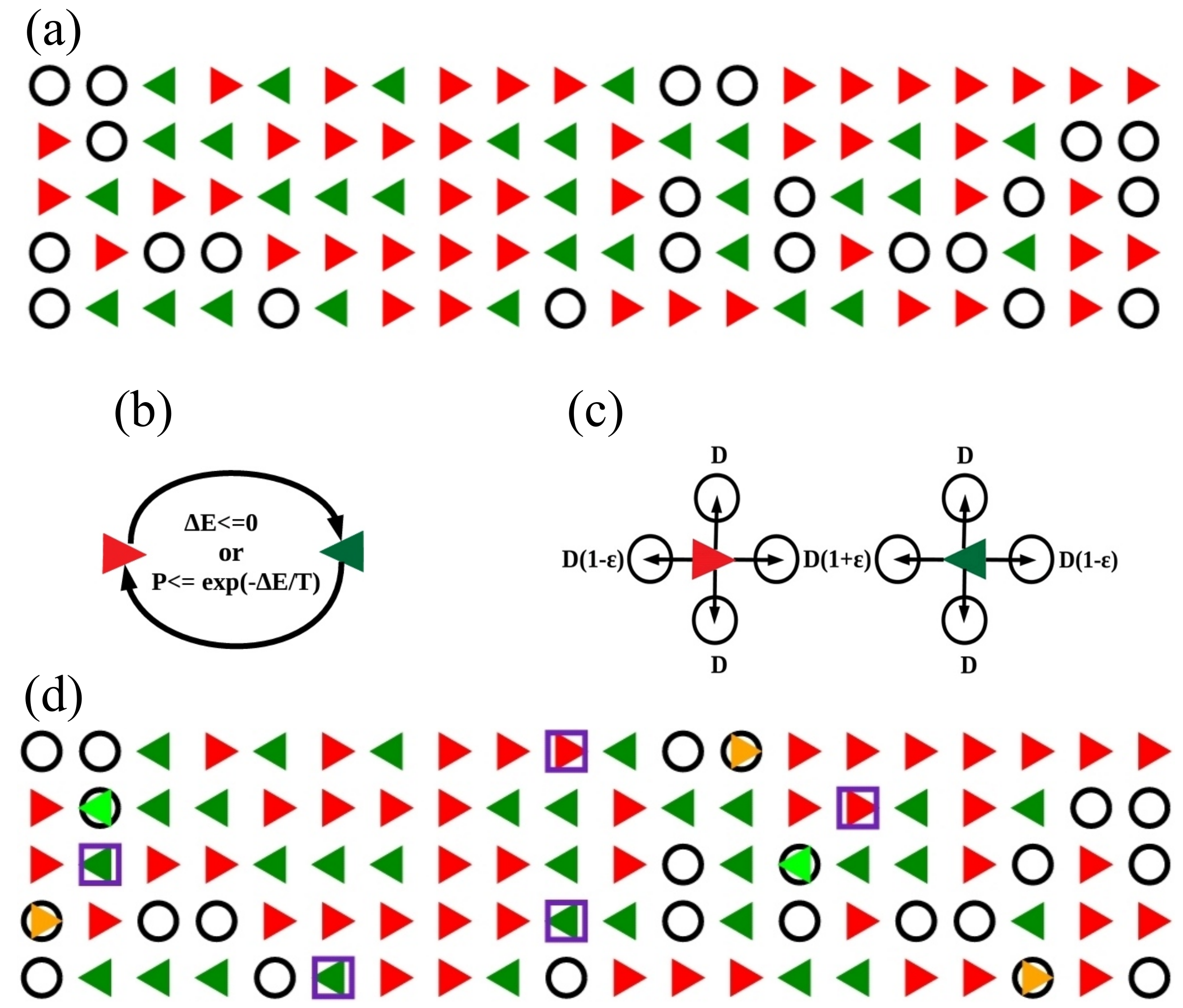}

	\caption{(color online) \textbf{(a)} Sketch of a part of system carrying spins $S=+1$(red (triangle right)) and $S=-1$(dark green (triangle left)) along with vacant sites S=0(black circle) on a two-dimensional lattice.\textbf{(b)} represents the flipping rate at fixed temperature.\textbf{(c)} represents the probability of movement of the spins to the neighboring sites. \textbf{(d)} birth and death rate $\gamma$ is added in the model in which the particles disappearing from random sites is shown by magenta square and appearing at the sites which were vacant represented by orange triangle right for $S=+1$ and green triangle left for $S=-1$. }

\label{fig1}
\end{figure}

\begin{figure}      
\includegraphics[width=8cm,height=6cm]{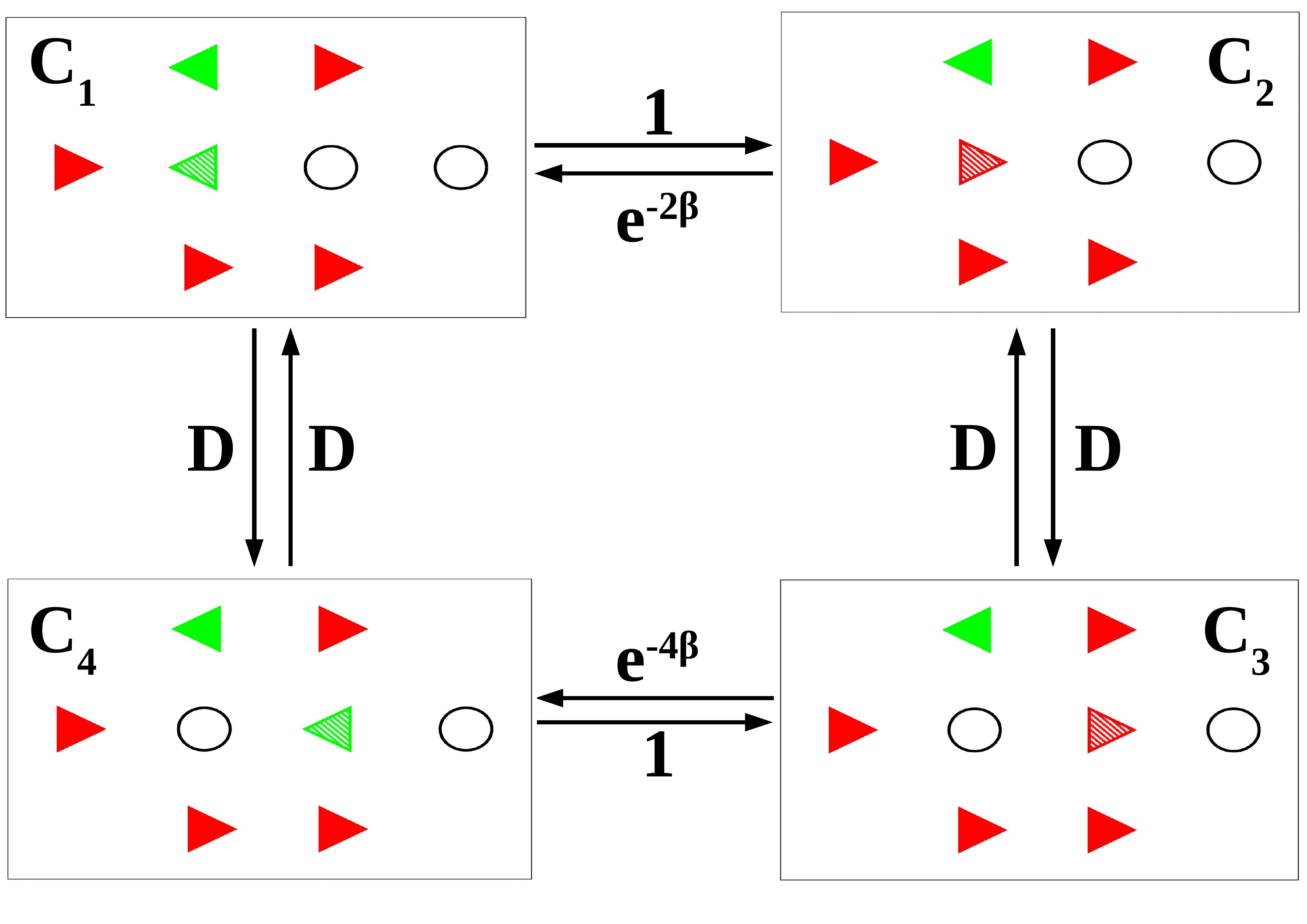} 
	\caption{(color online) A loop of four configurations that breaks Kolmogorov's criterion {\cite{kog}} of detailed balance for $dIm$, flipping and movement is shown for the spin represented by partially filled (red (triangle right))	and  (dark green (triangle left)). The clockwise loop $(C_1$  $\rightarrow$ $C_2$ $\rightarrow$ $C_3$ $\rightarrow$ $C_4$ $\rightarrow$ $C_1)$ gives the total probability $ D^2 e^{-4\beta}$ while the anticlockwise loop $(C_1$ $\rightarrow$ $C_4$ $\rightarrow$ $C_3$ $\rightarrow$ $C_2$ $\rightarrow$ $C_1) $ gives  the total probability $ D^2 e^{-2\beta}$, thus showing that the system does not satisfy detailed balance. The numbers associated to the arrows are the transition rates and other symbols have the same meaning as in Fig.{\ref{fig1}}.}
\label{fig: 2}
\end{figure} 

\section{Model and numerical details} 
We consider a system of active Ising spins $(AIM)$ 
on a two-dimensional rectangular lattice of size $L_x$ $\times$ $L_y$ 
with periodic boundary condition in both directions. 
A fraction of sites on the lattice is vacant.
 Each spin can take two possible values $S_i=\pm 1$.  Some of the 
 sites are vacant hence we define an occupancy variable 
$n_i=0$ or $1$ for the unoccupied and occupied sites respectively. Each site can have maximum one particle on it.
Hence, unlike the previous $AIM$ introduced by Solon {\em et. al.} our spins have mutual exclusion among them \cite{mutex}. 
Each spin can interact with its nearest neighbor spins using  the  Ising Hamiltonian \cite{ising}
\begin{equation}
	 H = - \sum_{i=1}^N n_i n_j S_i S_{j} , \label{eq:ising}
\end{equation}
hence the interaction term is non-zero only if the site and the interacting sites both are occupied. 
The above Hamiltonian Eq. \ref{eq:ising} is simulated  for a fixed  vacancy density $V=20\%$ (particle density $\rho=0.8$)
 by tuning the temperature. The temperature is introduced through the Metropolis Monte-Carlo algorithm  \cite{binder,pathria} for the 
alignment interaction among the spins. The ratio of the interaction strength and the Boltzmann constant  is chosen as $1$. 
The dynamics of the spins on the lattice can be  modeled in the following manner:  
(i)  the spins are fixed to their lattice sites $(fIm)$  and interacts through the Hamiltonian in Eq. \ref{eq:ising}. (ii) We allowed the spin to diffuse to any of its nearest vacant site
with equal probability. The model is named as diffusive Ising model $(dIm)$. 
In Fig. \ref{fig: 2} we show the Kolmogorov diagram \cite{kog} to check the detail balance condition 
on $dIm$. 
In Fig. \ref{fig: 2}, a loop of four configurations is shown, that breaks Kolmogorov's criterion, e.g. clockwise loop gives the total probability $ D^2 e^{-4\beta}$ while anticlockwise loop gives $ D^2 e^{-2\beta}$, thus showing that the system does not satisfy detailed balance. The numbers associated to the arrows are the transition rates. 
Hence $fIm$ satisfies the 
detail balance condition but the $dIm$ deviates from it.\\ 
(iii) Further we  made the spins active by introducing a biased movement 
corresponding to their direction as introduced in 
\cite{SolonPRE2015, SolonPRL2013}. The update rules showing the motion of the spins is shown in Fig. \ref{fig1}(b). Activity is introduced through a parameter $\epsilon \in (0,1)$. In the presence of activity $\epsilon$, the update rule for the movement
of the spins at a particular site is given as follows. 
 Each particle hops to its two neighboring sites left and right at rate $D(1 + S \varepsilon)$  
 provided the target site is vacant. It hops to other two sites (up and down) with equal probablity $D$.
If the $\varepsilon=0$, then the hoping rates are same in all the directions and that rate comes out to be 
$D=1/4$ and the model reduces to $dIm$. 
For nonzero $\epsilon$, the particle moves in the direction of its 
orientation at rate $D(1 + \varepsilon)$ and in the opposite direction to 
its orientation at rate $D(1 - \varepsilon)$. Whereas the particle hops with 
rate $D$ to other two possible directions. For our present study we fix $\varepsilon=1.0$.
We call the model
as active  model $(Am)$.  \\
(iv) Next we introduce  the birth and death of particles in the model 
$(Am)$.
The birth and death rate $\gamma$ is introduced as a fraction of sites on the lattice with density $\gamma$, such that from the 
randomly chosen $\gamma/2$ fraction of sites we  remove the particles (if occupied) and similarly by another randomly chosen $\gamma/2$, we  introduced the new particles with spin favoured with the majority spins in their nearest neighbor. 
The  $\gamma$ is tuned from $0$ to $0.1$. For $\gamma=0$ model reduces to $Am$ and for $\gamma \ne 0$  we call it birth and death active  model $(bdAm)$. 
One simulation step is counted after successful update of the above steps for all the particles once. Total simulation steps (time) used is $T=1.7\times10^5$.
The steady state in the system is achieved after
simulation time $7\times10^4$. We use $40$
independent realizations for averaging the data for the system size $L_x=400$ and $L_y= 50$.

\label{model}

\begin{figure}
\includegraphics[width=8cm,height=6cm]{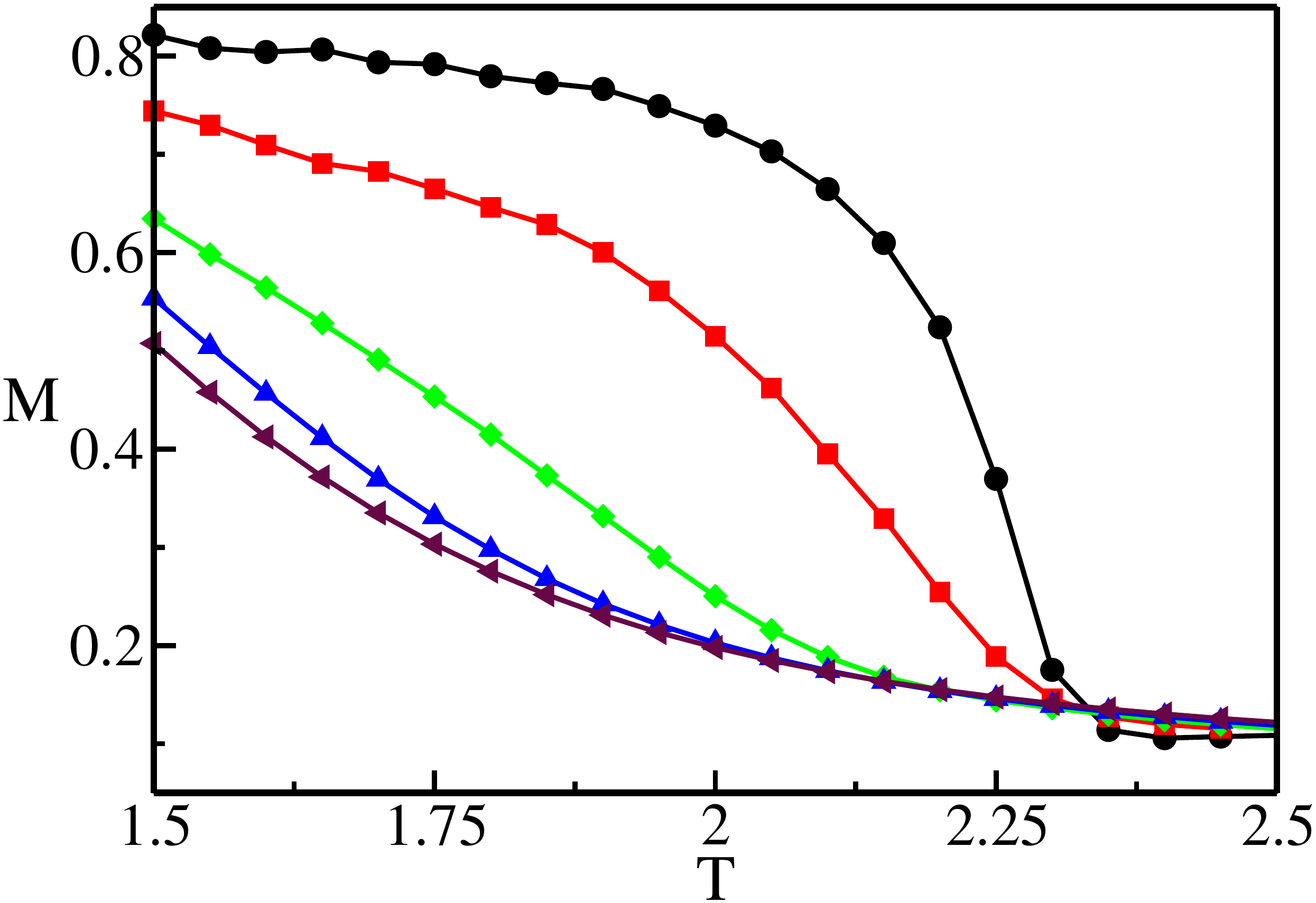}

\caption{The plot of order parameter $M$ vs. temperature $T$ for various values of $\gamma$ i.e. The black (circle), red  (square), green (diamond), blue (triangle right)  and maroon (triangle left) for  $\gamma = 0$, $\gamma= 0.01$, $\gamma = 0.1$, $\gamma = 0.5$ and $\gamma= 1$ respectively. The lines are guide to the eyes.}
\label{fig: 3}
\end{figure}

\begin{figure*}
\includegraphics[width=8cm,height=16cm, angle=-90]{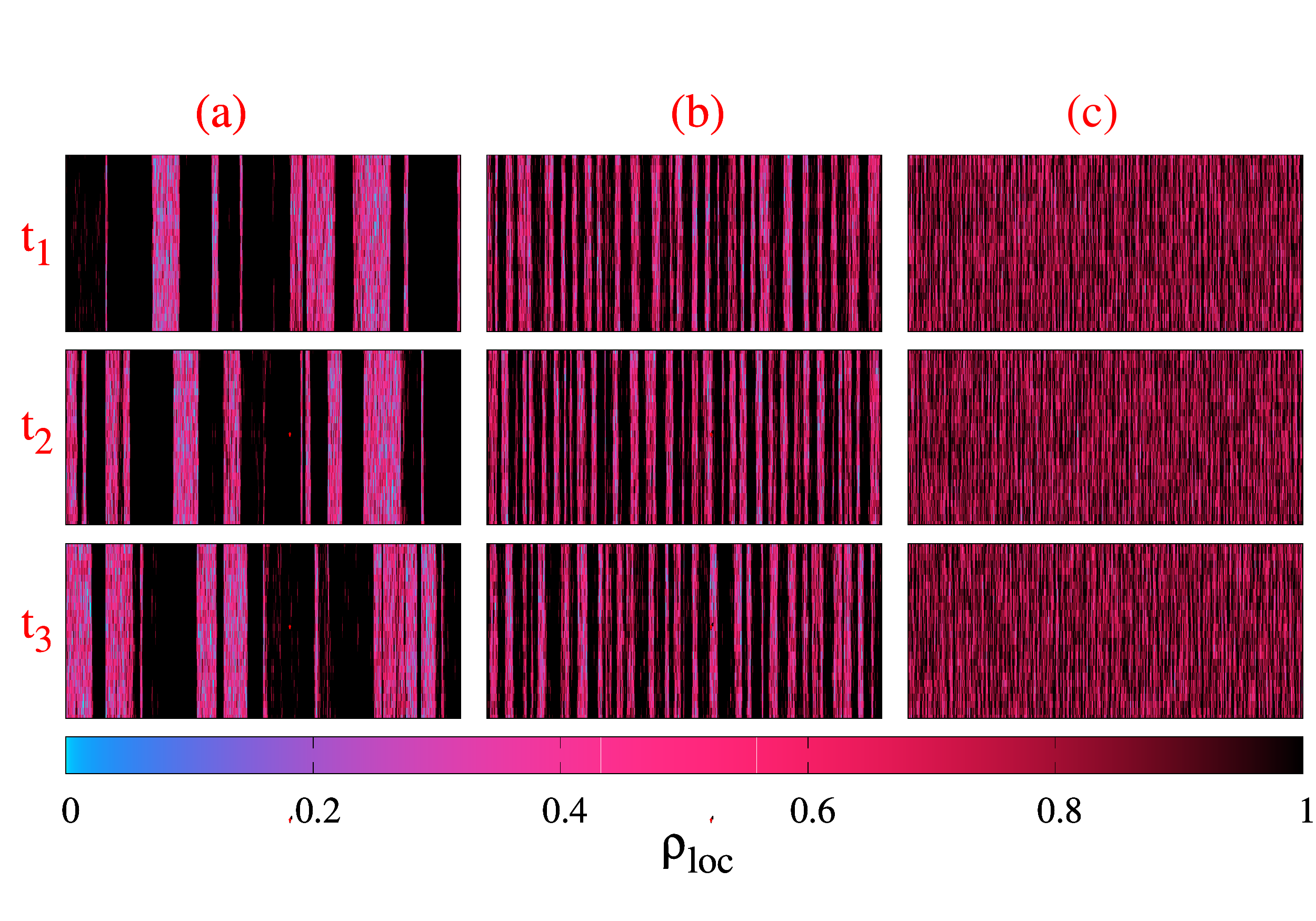}
\includegraphics[width=8cm,height=16cm, angle=-90]{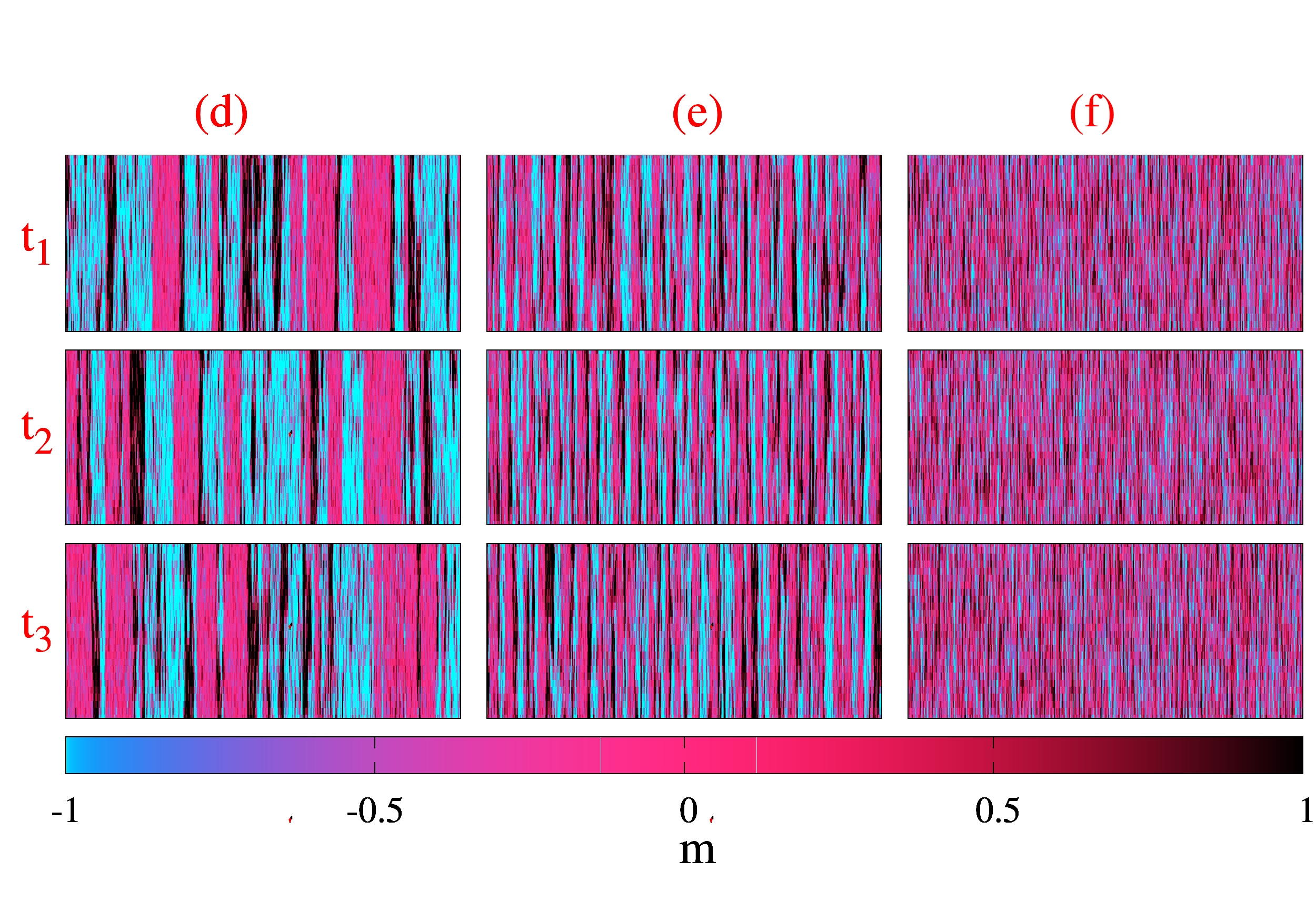}
	\caption{Time snapshots of local density $\rho_{loc}$ (a-c) and local magnetisation $m$ (d-f). From left to right (a-c) or (d-f) is for $\gamma =0.0$, $0.01$ and $0.1$ and ($t_1$-$t_3$) is for $1.4\times10^5$, $1.5\times10^5$, $1.6\times10^5$ respectively. Color bars represent the value of local density $\rho_{loc}$ and magnetisation $m$.}
\label{fig: 4}
\end{figure*}

\section{Results}
We first studied the   model $(bdAm)$ with 
no birth and death rate $\gamma=0$. The system is studied by varying 
temperature. A disorder-to-order phase transition is found on decreasing temperature. 
We studied the system for  activity $\epsilon=1$ and 
for different  $\gamma$. We first calculated  the global magnetisation in the system defined as:-

\begin{equation}
	M(t)  =  \frac{1}{N}\mid \sum_{i}  S_{i}(t) \mid 
\end{equation}
where $N$ is total number of particles. We define the mean magnetisation $M = < M(t)>$, where $< .. >$ means the average over time in the steady state and over different realisations.
We find that for the high temperature for all $\gamma$ system is disordered and $M \simeq 0$, and 
ordered with $M \simeq 1$ for low temperature. The variation of $M$ as 
a function temperature is shown in Fig. \ref{fig: 3} for different $\gamma$. We find a very
strong dependence  of  the shape of the disorder-to-order  curve on $\gamma$. The shape of the transition curve
changes from first order (discontinuous type) to continuous type as we tuned the 
$\gamma$ from $0$ to $0.1$. Hence, the nature of 
phase transition changes from first order type to continuous type for large  birth and death rate $\gamma$.\\
To further confirm
the nature of transition, we looked the system near  to the disorder-to-order transition.
We first plot  the real space snapshot of local density $\rho_{loc}$ in Fig. \ref{fig: 4}(a-c). 
The $\rho_{loc}$ is calculated by counting the density of spins in box of size $2 \times 2$. The panels from 
top to bottom ($t_1$ - $t_3$)  
for three different simulation times 
$t=1.4 \times 10^5$, $1.5 \times 10^5$ and 
$1.6 \times 10^5$ respectively. The panel (a)-(c) is for $\gamma=0$, $0.01$ and $0.1$ respectively.
For $\gamma=0$, we see the formation of bands of high density spins. 
With time the bands move across the system. The bands get diluted on increasing $\gamma$ 
and disappear for large $\gamma=0.1$.
The color bar shows the value of 
local density $\rho_{loc}$. 
Similarly we also plot the local magnetisation $m$, obtained by calculating the mean spin 
in the box of size $2 \times 2$ in Fig. \ref{fig: 4}(d-f) for the same set of parameters as for (a-c).
The panel (d-f) is for $\gamma=0$, $0.01$ and $0.1$ respectively.
We again find for zero $\gamma$, bands of high ordered region 
moves in the background of disordered region. The bands get diluted on increasing $\gamma$ and
finally disappear for large $\gamma=0.1$. The color bar shows the value of local $m$ and positive 
and negative $m$ represents the mean local spin $+1$ and $-1$ respectively. 
Very clearly the band splits into thinner and weaker bands on 
the introduction of  $\gamma$.
The formation of bands we find here for $\gamma=0$ or $Am$ is a common characteristics 
of polar flock \cite{chatepre2008, SolonPRE2015, SolonPRL2013, shradhapre2010}. 
For large $\gamma   \simeq 0.1$  slowly the density 
pattern disappears and its all become close to mean density $\rho_{loc}=0.8$.\\

\begin{figure}      
\includegraphics[width=8cm,height=6cm]{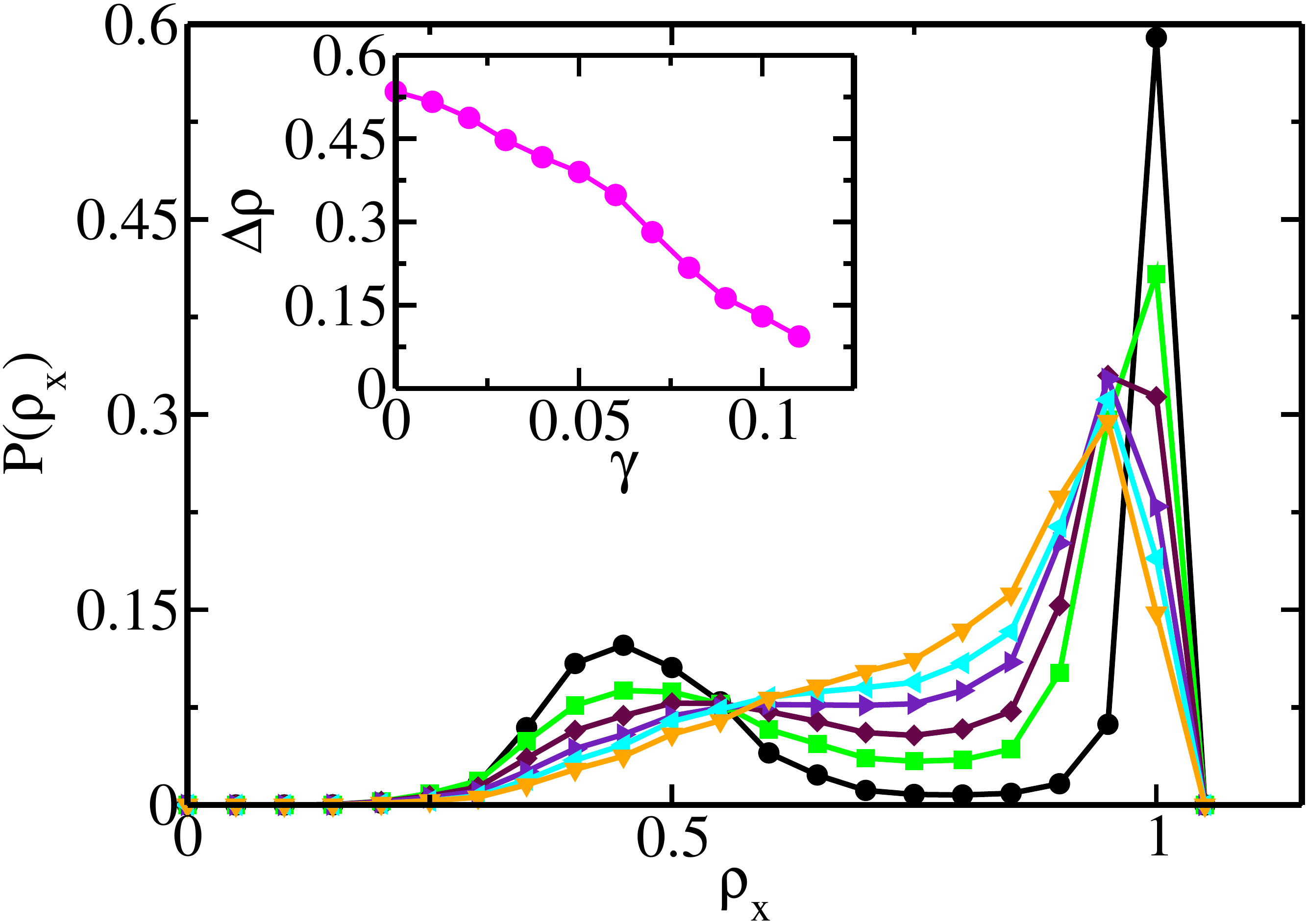} 
	\caption{(color online) Plot of PDF $P(\rho_x)$ for different values of $\gamma$  i.e. The black (circle), green (square), maroon (diamond), indigo (triangle right), cyan (triangle left) and orange (triangle down) symbols represent $\gamma = 0$, $\gamma= 0.02$, $\gamma = 0.04$, $\gamma = 0.06$, $\gamma = 0.08$ and $\gamma= 0.1$ respectively. The lines are guide to eyes. Inset: plot of $\Delta \rho$ vs. $\gamma$ where $\Delta \rho$ represents the difference between two peaks of the distribution of local density.} 
\label{fig: 5}
\end{figure}

\begin{figure}      
\includegraphics[width=8cm,height=7cm]{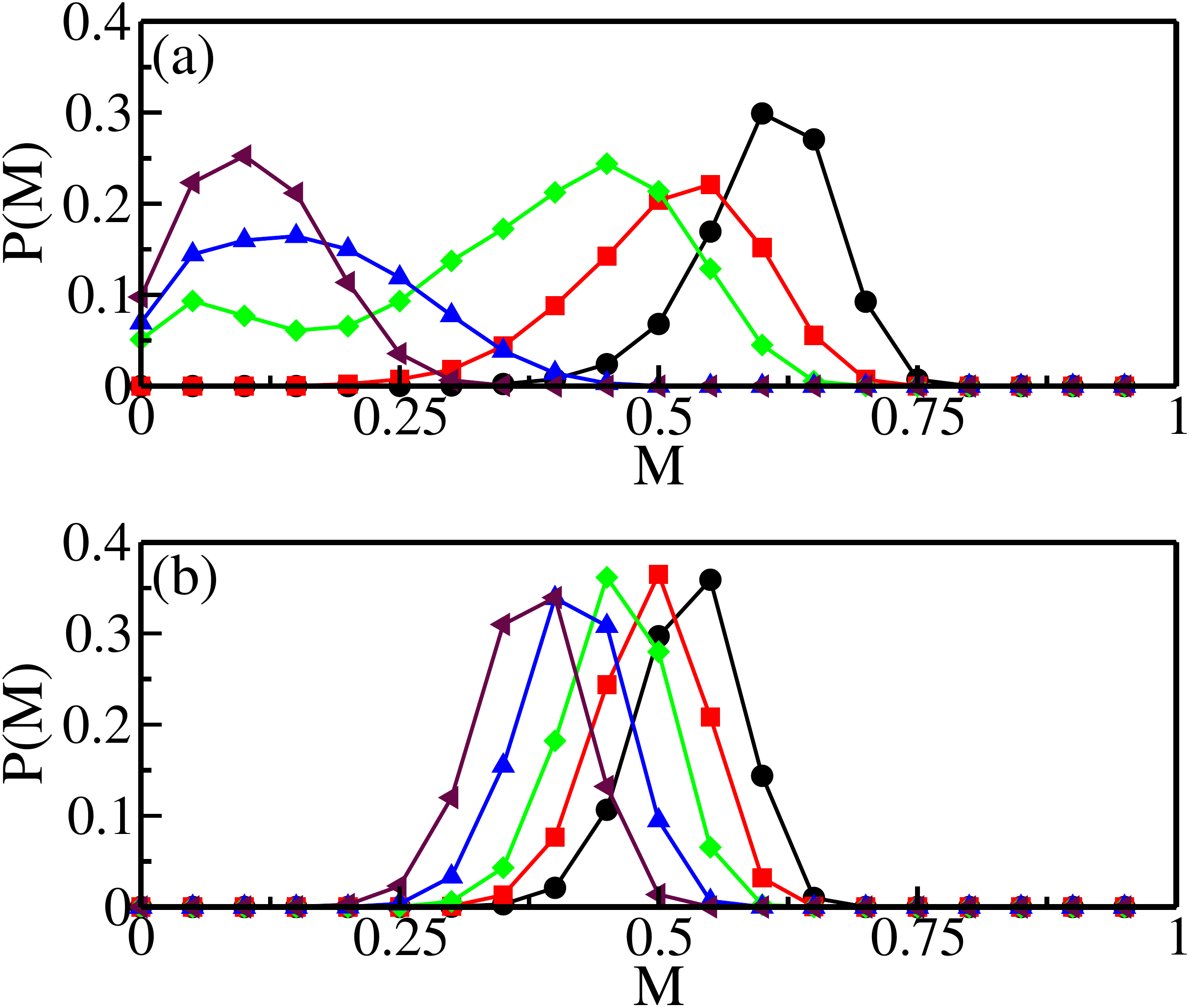} 
	\caption{(color online) Plot of PDF of order parameter $P(M)$ for $\gamma = 0 $ where the black (circle), red (square), green (diamond), blue (triangle left) and maroon (triangle up) symbols represents the  temperatures (a) $T=2.15$, $T=2.20$, $T=2.25$, $T=2.30$ and $T=2.35$ respectively and (b)    $T=1.65$, $T=1.70$, $T=1.75$ $T=1.80$ and $T=1.85$ respectively near the disorder-to-order transition. Lines are guide to eyes.}
\label{fig: 6}
\end{figure}

To further characterise the density inhomogeneity for different $\gamma$ we plot the distribution of density for different
temperatures close to disorder-to-order transition. 
Using the the local density $\rho_{loc}$ plots shown in Fig. \ref{fig: 4}(a-c),  we calculated the local density
along the long axis of the system by averaging over the shorter axis $L_y$. In this manner we find the density 
variation in one direction $\rho_x$. We further plot the
probability distribution function (PDF) of density $P(\rho_x)$ for various $\gamma$ in Fig. \ref{fig: 5}.
For zero $\gamma$, distribution clearly shows the bimodal nature, with one peak 
close to $1$ (maximum density) and another peak at lower density $\rho_x=0.4$.
As we increase $\gamma$ the two peaks come closer and finally for $\gamma \ge 0.1$ we find a single peak at $\rho_x=0.8$. In inset of Fig. \ref{fig: 5} we plot the density difference of two peaks $\Delta \rho$ vs. $\gamma$, and plot clearly shows a monotonic decrease of $\Delta \rho$ on increasing $\gamma$. \\
To understand the effect of $\gamma$  on the nature of phase transition in the system 
we observed time series of the global magnetisation
$M(t)$ in the steady state for two different $\gamma=0$ and $\gamma=0.1$. 
Using the time series we calculated the probability distribution function (PDF) of magnetisation $P(M)$.
In Fig. \ref{fig: 6} we plot the  $P(M)$ in the vicinity of disorder-to-order transition. 
 Fig. \ref{fig: 6}(a) is for $\gamma=0$ and for $\gamma=0.1$ is shown in 
 Fig. \ref{fig: 6}(b). Fig. \ref{fig: 6}(a), shows a bimodal distribution of $P(M)$ with one peak at $M=0.05$ and another at 
 $M=0.45$ for some intermediate temperature $T=2.25$ and in the neighborhood of
 $T=2.25$, $T = 2.15, 2.20, 2.30, 2.35$ we find jump in the peak position of $P(M)$.
 Whereas the distribution is always unimodal for all $T$ and the location of 
 peak in $P(M)$ smoothly moves towards lower $M$ values 
 for $\gamma=0.1$ as shown in Fig. \ref{fig: 6}(b).
Hence we say that the nature of the phase transition changes from the discontinuous to continuous type on increasing $\gamma$.
Now using coarse-grained hydrodynamic equations of motion we show how the increasing  birth and death term
in the density equation can lead to continuous transition.

\label{results}

\section{Coarse-grained hydrodynamic equations of motion}

Now we introduce the coarse-grained hydrodynamic equations of motion for slow variables: density $\rho({\bf r}, t)$ 
and polarisation
order parameter ${\bf P}({\bf r}, t)$. Former is globally conserved and later is nonzero for the broken symmetry 
state. The equation of motion for local density field $\rho({\bf r}, t)$ 
\begin{equation}
	\frac{\partial \rho}{\partial t} = - v_0 \nabla .({\bf P} \rho)+ D_{\rho} \nabla^2 \rho + \gamma g({\rho})
	\label{eqrho}
\end{equation}
The hydrodynamic equation of motion for the local polarisation ${\bf P}({\bf r}, t)$, 
\begin{equation}
	\frac{\partial {\bf P}}{\partial t} = (\alpha_1 (\rho)-\alpha_2 \mid {\bf P}\mid ^2){\bf P} - v_1 \nabla \rho +\lambda ({\bf P} \cdot \nabla{\bf P})+D\nabla^2 {\bf P}
    \label{eqp}
\end{equation}\\
Equation \eqref{eqrho} represents continuity equation with additional birth and death term $g(\rho)= \rho(\rho-\rho_0)$ for the  density field $\rho$. Where $\rho_0$ is the mean density of particles in the system. 
$\gamma$ is the birth and death rate. 
The first term on the right hand side of Eq. \ref{eqrho}, describes the convection due to self-propulsion velocity $v_0 \bf P$. 
The second term on the right hand side
of Eq. \ref{eqrho} is diffusion due to density gradient.
In polarisation ${\bf P}({\bf r}, t)$, Eq. \eqref{eqp}, the first term on right hand side represents a mean field transition from an
isotropic state ($\bf P = 0$) to a broken symmetry state ${\bf P} =\sqrt{\frac{\alpha_1(\rho_0)}{\alpha_2}} {\bf \hat{x}}$
(the direction of broken
symmetry is chosen along $x-$axis). The second and third term indicate hydrostatic pressure
due to density gradient and convection in the model, respectively. Both $\lambda$ and $v_1$ depends
on self-propelled speed of the particle \cite{bertin2006}. The
fourth term represents diffusion in the polarisation field.
The above two equations are similar to the equations introduced in \cite{tonertupre1998}. Here we introduced an
additional term due to birth and death in the density equation.
We first analyse the equations for the broken symmetry or the ordered homogeneous state of the two Eqs. \ref{eqrho} and \ref{eqp}, $\rho=\rho_0$ and 
${\bf P}=P_0=\sqrt{\frac{\alpha_1(\rho_0)}{\alpha_2}}\hat{\bf x}$.
 We further add  small
perturbation on the above homogeneous ordered state and write:
$\rho =\rho_0 + \delta \rho({\bf r}, t)$ and ${\bf P} =(P_0 + \delta P_x)\hat{x}+\delta P_y\hat{y} $.
 $\delta \rho({\bf r}, t)$, $\delta P_x({\bf r}, t)$ and $\delta P_y({\bf r}, t)$ are the fluctuations in the density, longitudinal and transverse directions of polarisation respectively. 
Since system shows a mean-field transition from disordered-to-ordered state where $\alpha_1$, changes sign. Hence at the transition point
$\alpha_1=0$. We take $\alpha_1=0$ (at the mean-field transition point) and  substitute for the $\rho$ and ${\bf P}$ from the above expressions 
and further write the Eqs. \ref{eqrho} and \ref{eqp} for the 
small fluctuations $\delta \rho, \delta P_x, \delta P_y$ to the linear order. 
The equation for $\delta P_x$, will not contribute to linear order. Hence only the equations for the local density fluctuation $\delta \rho({\bf r}, t)$ and 
local transverse polarisation fluctuation $\delta P_y({\bf r}, t)$ will survive. We further take the Fourier transform of the two equations using 
${\bf A}({\bf q}, \omega) = \int{{\bf A}({\bf r}, t) \exp(i {\bf q} \cdot {\bf r} + i \omega t)  d {\bf r} d \omega}$, where ${\bf A} = (\delta \rho, \delta P_y)$. Hence the two equations will become:

\begin{equation}
	(-i \omega + D_{\rho} q^2 + \gamma\rho_0)\delta \rho + i q_y v_0 \rho_0 \delta P_y=0
	\label{eqdelta}
\end{equation}
and 
\begin{equation}
	i  q_y \frac{v_1}{2\rho_0} \delta \rho + (-i \omega + D q^2)\delta P_y=0
	\label{eqdeltap}
\end{equation}
where the wavevector $q$ is in the direction of broken symmetry. We further 
do the analysis for the transverse direction $q_y=q$.
We further write the above two Eqs. \ref{eqdelta} and \ref{eqdeltap} in matrix notation and solve for the two modes $\omega_{\pm}$ using 
the determinant of the matrix and find the two modes as
\begin{equation}
	\omega_{\pm}=\frac{-i\gamma \rho_0}{2}\pm\frac{iq}{2}\sqrt{2 v_0 v_1 +2\gamma\rho_0(D_\rho-D)}
	\label{eqomega}
\end{equation}\\
We further use the solution for the two modes to performed the renormalised mean-field study of the system in the isotropic state. 
In the next section we carry out the  perturbative study of the hydrodynamic equations about the isotropic state.
\label{analytics}

\subsection{Perturbative renormalised mean-field study  in the isotropic state}
We additionally introduce small
fluctuations about the isotropic  state;  $\rho=\rho_0+\delta \rho$ and
${\bf P}=(\delta P_x, \delta P_y)$ and write the equations for the small
fluctuations $\delta \rho$ and $\delta {\bf P} = (\delta P_x, \delta P_y)$. The  solution for the Fourier 
transformed local density fluctuation $\delta \rho({\bf q}, \omega)$ and polarisation fluctuations $\delta {\bf P}({\bf q}, \omega)$ will become;
\begin{widetext}
\begin{equation}
 \begin{aligned}
	 \delta \rho({\bf q}, \omega)={} & \ \frac{-v_0\rho_0 i {\bf q}.{\delta{\bf P}({\bf q}, \omega)}}{(-i\omega+D_\rho q^2 +\gamma\rho_0)}\\ & \ -\frac{1}{2}\frac{v_0i{\bf q}.\displaystyle\int{[\delta P({\bf k}, \Omega)\delta \rho({{\bf q}-{\bf k}}, \omega-\Omega)+\delta P({{\bf q}-{\bf k}}, \omega-\Omega) \delta\rho(k,\omega)]d{\bf k} d\Omega}}{(-i\omega+D_\rho q^2 +\gamma\rho_0)}\\ 
	\end{aligned}
	\label{eq8}
\end{equation}
\end{widetext}
and \\
\begin{widetext}
\begin{equation}
\begin{aligned}
	-i\omega \delta {\bf P}({\bf q},\omega)={} & \-\alpha_1(\rho_0)\delta {\bf P}({\bf q}, \omega) \\ & \ +{\alpha_1}'(\rho_0)\frac{1}{2}\left[\int{\delta {\bf P}({\bf k}, \Omega) \delta \rho({\bf q}-{\bf k}, \omega-\Omega) + \delta {\bf P}({\bf q}-{\bf k}}, \omega-\Omega) \delta \rho({\bf k}, \Omega) d{\bf k} d\Omega\right]\\& \ -\alpha_2 \int{\delta {\bf P}({\bf k}, \Omega) \delta {\bf P}({\bf k'}, \Omega-\Omega') \delta {\bf P}({\bf q}-{\bf k}-{\bf k}', \omega-\Omega-\Omega') d{\bf k} d{\bf k'} d\Omega d \Omega}
	\end{aligned}
	\label{eq9}
\end{equation}
\end{widetext}
Here we write only first two terms in the polarisation Eq. \ref{eqp}. $\alpha_1'(\rho_0)=\frac{\partial{\alpha_1}}{\partial\rho}\Big\vert_{\rho=\rho_0}$.
Substitute for $\delta \rho$ leading order from Eq. \ref{eq8} and substituting for one of mode $\omega_{+}$ from Eq. \ref{eqomega}
\begin{widetext}
\begin{equation}
\begin{aligned}
	-i\omega \delta {\bf P}({\bf q},\omega)={} & \ \alpha_1(\rho_0)\delta {\bf P}({\bf q}, \omega) \\ & \ +\frac{1}{2}{\alpha_1}'(\rho_0)\Bigg[\frac{\int(\delta {\bf P}({\bf k})(v_0 \rho_0({\bf q}-{\bf k}))\delta {\bf P}({\bf q}-{\bf k}, \omega-\Omega))}{6\gamma\rho_0+(q- k)\sqrt{2 v_0 v_1 +\gamma\rho_0(D_\rho-D)}} \\ & \ +  \frac{\int(\delta {\bf P}({\bf q}-{\bf k},\omega-\Omega)(v_0 \rho_0{\bf k})\delta P({\bf k},\Omega))}{\gamma \rho_0+k\sqrt{2 v_0 v_1 +\gamma\rho_0(D_\rho-D)}}\Bigg]  \\& \ -\alpha_2 \int{\delta {\bf P}({\bf k}, \Omega) \delta {\bf P}({\bf k'}, \Omega-\Omega') \delta {\bf P}({\bf q}-{\bf k}-{\bf k}', \omega-\Omega-\Omega') d{\bf k} d{\bf k'} d\Omega d \Omega}
\end{aligned}
\label{eq10}
\end{equation}
\end{widetext}
It is  complicated to solve the above equation for the  dimensions $d = 2$, when $\delta {\bf P}$ is a vector. However the usual flocking transition is characterised 
by the appearance of bands near the transition \cite{chatepre2008, SolonPRL2013, shradhamanna2016}. 
When bands form, the local density and polarisation shows the variation only along the direction of moving
bands and in the transverse direction it is homogeneous both in space and time. Hence system can be considered one dimensional where both
$\delta {\bf P}$ and $\delta \rho$, only vary along the direction of moving bands. And all the vectors can be replaced by scalars in Eqs. \ref{eq10}.
In such conditions, we can rewrite the above Eq. \ref{eq10} as 
\begin{widetext}
\begin{equation}
\begin{aligned}
	-i\omega \delta  P({\bf q},\omega)={} & \ \alpha_1(\rho_0) \delta P(q,\omega) \\ & \ + \frac{1}{2}\alpha_1'(\rho_0) \bigg[\frac{v_0 \rho_0 \int{\delta P (q-k,\omega-\Omega) (q-k)\delta P(k,\Omega) dk d\Omega}}{6 \gamma \rho_0 + (q-k) \sqrt{2 v_0 v_1 + \gamma \rho_0(D_{\rho}-D)}} \\ & \ + \frac{v_0 \rho_0 \int{\delta P (q-k,\omega-\Omega) k \delta P(k,\Omega) dk d\Omega}}{6 \gamma \rho_0 + k \sqrt{2 v_0 v_1 + \gamma \rho_0(D_{\rho}-D)}} \bigg] \\ & \ - \alpha_2 \int{\delta P(k, \Omega) \delta P(k', \Omega') \delta P(q-k-k', \omega-\Omega-\Omega') dk dk' d\Omega d\Omega'}
\end{aligned}
	\label{eq11}
\end{equation}
\end{widetext}
{\it {\bf Case with no birth and death ($\gamma=0$)}}:\\
Now for the zero $\gamma$ or no birth and death, we write the effective free energy $\mathcal{F}_{eff}(\delta P)$ using Eq. \ref{eq11} as
\begin{widetext}
\begin{equation}
	\begin{aligned}
\mathcal{F}_{eff}(\delta P)={} & \ -\alpha_1(\rho_0)\int{dk\frac{\delta P(k) \delta P(q-k)}{2}}  \\
	& \ - \frac{\alpha_1'(\rho_0)}{6}\sqrt{\frac{v_0}{2 v_1}}\int{dk dk'\left[\delta P(k) \delta P(q-k-k')\delta P(k')\right]}  \\ 
	& \ +\alpha_2 \frac{1}{4}\int{dk dk' dk'' \delta P(k) \delta P(k') \delta P(k'') \delta P(q-k-k'-k'')}
      \end{aligned}
\end{equation}
\end{widetext}
Taking the inverse Fourier transform, the expression for the  $\mathcal{F}_{eff}({\delta P})$ and assuming the homogeneous $\delta P$. The effective free energy $\mathcal{F}_{eff}(\delta P)$
is real space will become
\begin{widetext}
\begin{equation}
\begin{aligned}
	\mathcal{F}_{eff}(\delta P)=-\alpha_1(\rho_0)\frac{\delta P^2}{2}- \frac{\alpha_1'(\rho_0)}{6}\sqrt{\frac{v_0}{2 v_1}}\delta P^3 +\frac{\alpha_2}{4}\delta P^4
\end{aligned}
	\label{eqfreereal}
\end{equation}
\end{widetext}
Hence for the zero birth and  death, the second term of the right hand side is an additional expression which
is cubic order in $\mathcal{O}(\delta P^3)$. The presence of such nonlinear term can leads the mean-field transition to first order. 
Now we examine the system for finite $\gamma$, using the denominator of Eq. \ref{eq11}
\begin{equation}
\gamma \rho_0 \gg \frac{q\sqrt{2v_0 v_1+\gamma\rho_0(D_\rho - D)}}{6}
\label{eq15}
\end{equation}\\
Then, the second term in Eq. \ref{eq10} will be  of the form  $P \nabla P$ and this term can be compared with 
the  convective nonlinear term of the type ${\bf P} \cdot \nabla {\bf P}$ in the hydrodynamic Eq. \ref{eqp}.
After solving Eq. \ref{eq15} for $\gamma$ we get,
\begin{equation}
	\gamma > \frac{q \Delta D}{72 \rho_0}(1 + \sqrt{q^2 + \frac{288 v_0 v_1}{\Delta D}^2})
	\label{eq16}
\end{equation}

        
where $\Delta D=|D_\rho - D|$. Hence if $\gamma$ is greater than the right hand side of Eq. \ref{eq16},  the birth and death term  
will dominate and transition will be of type as predicted by mean field theory, whereas for finite and small $\gamma$ such that
\begin{equation}
\gamma \rho_0 \ll \frac{q\sqrt{2v_0 v_1+\gamma\rho_0(D_\rho - D)}}{6}
\end{equation}
the effective free energy for $\mathcal{F}_{eff}(\delta P)$ will  become
\begin{widetext}
\begin{equation}
\begin{aligned}
	F_{eff}(\delta P)={}&\ -\frac{1}{2}\alpha_1(\rho_0)\delta P^2 +\\
	& \ -\frac{v_0 \rho_0\alpha_1 '(\rho_0) \delta P^3}{6 \sqrt{2 v_0 v_1 +\gamma \rho_0(D_\rho-D)}}  +\frac{\alpha_2}{4} \delta P^3
\end{aligned}
\end{equation}
\end{widetext}
Hence for small $\gamma$, the wavevector dependence of  additional convective non-linear term in Eq. \ref{eq11} goes  away
and it contributes an additional $\mathcal{O}(\delta P^3)$ nonlinearity in the effective free energy. Which will lead the transition
to discontinuous type or  first order. Hence the first order or discontinuous transition happens through the competition of 
a length  scale and birth and death rate as given in Eq. \ref{eq16}. For large wavevector $q$ (or small wavelength) term on the right hand side of
Eq. \ref{eq16}, larger $\gamma$ will make the transition continuous type and vice versa. Hence the wavelength ($q^{-1}$) of the density and magnetisation
fluctuations decreases with increasing birth and death term.
As shown in Fig. \ref{fig: 4},
on increasing $\gamma$ from $\gamma=0$, the bands start to split and their  size deceases, and the nature of transition becomes more
and more continuous type as predicted by mean-field type.\\

We further analysed the properties of effective free energy in the presence of additional cubic order nonlinearity.
In simplified notation the effective free energy can be written as 
\begin{equation}
	\mathcal{F}(\delta P)=-\beta_1 \delta P^2 -\beta_2 \delta P^3 +\beta_3 \delta P^4
\end{equation}
where
$\beta_1=\frac{1}{2}\alpha_1(\rho_0)$, $\beta_2=\frac{v_0\rho_0\alpha_1 '(\rho_0)}{6 \sqrt{2 v_0 v_1 +\gamma \rho_0(D_\rho-D)}}$, $\beta_3=\frac{\alpha_2}{4}$.
For the transition to be first order we impose the coexistence condition  i.e. $\mathcal{F}(\delta P=0)=\mathcal{F}(\delta P\neq 0)$, that gives
\begin{equation}
-\beta_1 -\beta_2 \delta P +\beta_3 \delta P^2=0
\label{eq20}
\end{equation}
also the condition of  steady state implies $\frac{\partial \mathcal{F}}{\partial \delta P}=0$
\begin{equation}
-2 \beta_1-3 \beta_2 \delta P +4 \beta_3 \delta P^2=0
\label{eq21}
\end{equation}
using Eq. \ref{eq20} and \ref{eq21} the jump in the order parameter $P$ at the transition
\begin{equation}
	\delta P=\frac{\mid\beta_2\mid}{2\beta_3} 
\end{equation}
and the jump is always positive, hence the finite jump. Putting the value of $\delta P$ in Eq. \ref{eq20} and solve for $\beta_1$,
$\beta_1-\frac{\beta_2 ^2}{4 \beta_3}=0$. We get $\beta_1 ^c =\frac{\beta_2 ^2}{4 \beta_3}$  again a positive term. 
We further analyse the jump in the order parameter and transition point using
$\beta_2=\frac{v_0 \rho_0 \alpha_1 '(\rho_0)}{6\sqrt{2 v_0 v_1+\gamma \rho_0 \vartriangle D}}$
and  $ \delta P = \frac{4 v_0 \rho_0 \alpha_1 '(\rho_0)}{3 \alpha_3\sqrt{2 v_0 v_1 +\gamma\rho_0 \vartriangle D}}$. Assuming the temperature dependence of $\alpha_1$ in the mean-field 
theory
$\alpha_1(\rho, T)=\alpha_0(\rho)(T-T^*)$, $T^*$ is point where there is a mean-field type second order phase transition for large $\gamma$.
Hence if we define the true critical temperature as $T_c$, where $\beta_1= \frac{\beta_2^2}{4 \beta_3}$, and at the critical point $\alpha_1^c(\rho, T)  =\alpha_0(\rho)(T_c-T^*) = 2 \beta_1$ $\> 0$. Hence we find that  $T_c > T^*$ and 
$T_c=\frac{\beta_2^2}{2 \beta_3 \alpha_0} + T^*$. After substituting the value of $\beta_2$ and $\beta_3$ from the previous expressions, we find 
$T_c=T^*+\frac{2 {v_0}^2 {\rho_0}^2 {{\alpha_0}'}^2 \rho_0}{9 (2 v_0 v_1 +\gamma \rho_0 \vartriangle D)\alpha_2\alpha_0(\rho_0)}$. Hence in the simplified notation we can write
$T_c=T^* + \frac{A}{B+\gamma}$
where $A=\frac{4 {v_0}^2 {\rho_0}^2 {{\alpha_0}'} ^2\rho_0}{9\rho_0 \Delta D \alpha_3 \alpha_0(\rho_0)}$ and $B=\frac{2v_0 v_1}{\rho_0 \Delta D}$. For  large $\gamma$, second term in the expression for $T_c$  is negligible and critical point happens at $T^*$. As we start tuning $\gamma$ towards lower values, the phase transition shift towards right as obtained in our numerical study Fig. \ref{fig: 2}. 
Also $\delta P$, (the jump in order parameter) is almost zero for large $\gamma$ that means system approaches critical point continuously as found in our numerical study Fig. \ref{fig: 6}(b). 
But as we start decreasing $\gamma$ transition happens with a finite jump in order parameter Fig. \ref{fig: 6}(a).
 
\section{Discussion}
We studied a system of active Ising spins with the presence of birth and death on a two dimensional substrate with periodic boundary condition. The system is studied using Metropolis-Monte -Carlo for the interaction among the spins and the spin perform the biased move along their direction of orientation. System is studied for fixed activity and varying birth and death rate $\gamma$. Its shows a phase transition from disorder-to-order for all $\gamma$ and fixed activity on varying temperature from  high to low. The transition is of first order discontinuous type for conserved model ($\gamma=0$) and becomes continuous type for the birth and death model. Also transition shifts towards higher temperature on decreasing $\gamma$. Hence the presence of birth and death rate  tune the disorder-to-order transition to lower temperature  and shows a crossover from discontinuous to continuous type in the polar flock. The results are verified with the help of coarse-grained hydrodynamic equations of motion for local density and polarisation in the presence of birth and death.\\
Hence our study shows effect of birth and death on the nature of phase transition of polar-flock. The present model is studied for discrete Ising spins with the Globular conserved model \cite{binder1} 
for the spin 
interaction. It is worth to study the system for the non-conserved Kawasaki type \cite{binder2} of spin interaction as well for the off-lattice systems.       
\label{discussion}

{\em Acknowledgement : }  SM, thanks J K Bhattacharjee for useful discussion at the start of the project. SM also thanks S. Ramaswamy and M. C. Marchetti for introducing 
the problem a few years back. SM and PKM, thanks PARAM Shivay for computatational facility under the National Supercomputing Mission, Government of India at the Indian Institute of Technology, Varanasi. Computing facility at Indian Institute of Technology(BHU), Varanasi is gratefully acknowledged.

\bibliographystyle{apsrev4-1}
\bibliography{references} 

\begin{thebibliography}{33}
\bibitem [{mut()}]{mutex}%
  \BibitemOpen
  \href@noop {} {\bibinfo  {journal} {The mutual exclusion help us to introduce
  birth and death in simple manner (which we explain later)}\ }\BibitemShut
  {NoStop}%
\end{thebibliography}

\end{document}